\documentclass[acmtog, authorversion, nonacm]{acmart}
\acmSubmissionID{255}

\usepackage{graphicx} % allow us to embed graphics files
\usepackage{booktabs} % For formal tables
\usepackage{soul}                    % Color highlighting.
\usepackage{multirow}

\usepackage{pifont}
\newcommand{\cmark}{\ding{51}}%
\newcommand{\xmark}{\ding{55}}%

\newcommand{\etal}{~et~al.\ }

\newcommand{\mycomment}[1]{}

\definecolor{darkred}{rgb}{0.6,0,0}
\definecolor{green}{rgb}{0.0,0.5,0}
\definecolor{blue}{rgb}{0,0,0.75}
\definecolor{orange}{rgb}{1,0.6,0.2}
\definecolor{red}{rgb}{1,0,0}
\definecolor{purplish}{rgb}{0.6,0,0.7}

\def\real{\mathbb{R}}
\def\cff{\Psi}
\def\fs{f_{s}}
\def\ft{f_{t}}
\def\ecc{e}
\def\lum{l}
\def\lumcd{L}
\def\radius{u}
\def\acuity{A}

\def\pt{f_t} % measured point
\def\extent{\mathbf{m}}

\graphicspath{{"figures/"}}

\usepackage[ruled]{algorithm2e} % For algorithms

\SetAlFnt{\small}
\SetAlCapFnt{\small}
\SetAlCapNameFnt{\small}
\SetAlCapHSkip{0pt}

\acmJournal{TOG}
\usepackage{array}
\newcolumntype{P}[1]{>{\centering\arraybackslash}p{#1}}

\begin{document}
% Title portion
%\title{Towards Spatio-temporal Foveated Video Compression}
\title{A Perceptual Model for Eccentricity-dependent Spatio-temporal Flicker Fusion and its Applications to Foveated Graphics}

% DO NOT ENTER AUTHOR INFORMATION FOR ANONYMOUS TECHNICAL PAPER SUBMISSIONS TO SIGGRAPH 2019!
\author{Brooke Krajancich}
\affiliation{%
  \institution{Stanford University}
}
\email{brookek@stanford.edu}
\author{Petr Kellnhofer}
\affiliation{%
  \institution{Stanford University and Raxium}  
}
\email{pkellnho@stanford.edu}
\author{Gordon Wetzstein}
\affiliation{%
 \institution{Stanford University}
}
\email{gordon.wetzstein@stanford.edu}

\renewcommand{\shortauthors}{Krajancich, et al.}

\begin{abstract}
Virtual and augmented reality (VR/AR) displays strive to provide a resolution, framerate and field of view that matches the perceptual capabilities of the human visual system, all while constrained by limited compute budgets and transmission bandwidths of wearable computing systems. Foveated graphics techniques have emerged that could achieve these goals by exploiting the falloff of spatial acuity in the periphery of the visual field. However, considerably less attention has been given to temporal aspects of human vision, which also vary across the retina. This is in part due to limitations of current eccentricity-dependent models of the visual system. We introduce a new model, experimentally measuring and computationally fitting eccentricity-dependent critical flicker fusion thresholds jointly for both space and time. In this way, our model is unique in enabling the prediction of temporal information that is imperceptible for a certain spatial frequency, eccentricity, and range of luminance levels. We validate our model with an image quality user study, and use it to predict potential bandwidth savings 7$\times$ higher than those afforded by current spatial-only foveated models. As such, this work forms the enabling foundation for new temporally foveated graphics techniques.
\end{abstract}





\begin{teaserfigure}
   \includegraphics[width=\textwidth]{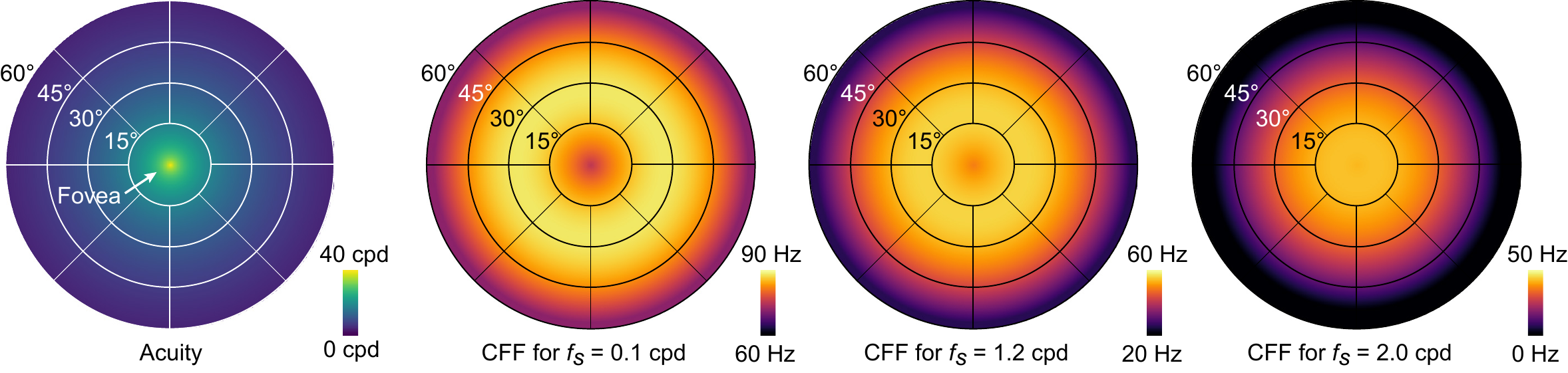}
   \caption{Foveated graphics techniques rely on eccentricity-dependent models of human vision. While such models are well understood for spatial acuity (left, \cite{geisler_real-time_1998}), our work is the first to experimentally derive a more comprehensive model for the spatio-temporal aspects over the retina under conditions close to virtual reality applications. As seen in the three plots on the right, we model critical flicker fusion (CFF) thresholds in an eccentricity-dependent manner. The CFF varies with spatial frequency $\fs$ and luminance and it exhibits an anti-foveated effect, with the highest thresholds observed in the near--mid periphery of the visual field. Our perceptual model and its experimental validation could provide the foundation of future spatio-temporally foveated graphics systems.}	
\end{teaserfigure}

\maketitle

{
\renewcommand{\thefootnote}{\fnsymbol{footnote}}
\footnotetext[1]{\href{https://www.computationalimaging.org/publications/cff/}{Project website: \\ https://www.computationalimaging.org/publications/cff/}}
}

%%%%%%%%%%%%%%%%%%%%%%%%%%%%%%%%%%%%%%%%%%%%%%%%%%%%%%%%%%%%%%%%%%%%
%%%%%%%%%%%%%%%%%%%%%%%%%%%%%%%%%%%%%%%%%%%%%%%%%%%%%%%%%%%%%%%%%%%%
% body

\section{Introduction}
\label{sec:intro}
Emerging virtual and augmented reality (VR/AR) systems have unlocked new user experiences by offering unprecedented levels of immersion. With the goal of showing digital content that is indistinguishable from the real world, VR/AR displays strive to match the perceptual limits of human vision. That is, a resolution, framerate, and field of view (FOV) that matches what the human eye can perceive. However, the required bandwidths for graphics processing units to render such high-resolution and high-framerate content in interactive applications, for networked systems to stream, and for displays and their interfaces to transmit and present this data is far from achievable with current hardware or standards.

Foveated graphics techniques have emerged as one of the most promising solutions to overcome these challenges. These approaches use the fact that the acuity of human vision is eccentricity dependent, i.e., acuity is highest in the fovea and drops quickly towards the periphery of the visual field. In a VR/AR system, this can be exploited using gaze-contingent rendering, shading, compression, or display (see Sec.~\ref{sec:related}). While all of these methods build on the insight that acuity, or spatial resolution, varies across the retina, common wisdom also suggests that so does temporal resolution. In fact it may be anti-foveated in that it is higher in the periphery of the visual field than in the fovea. This suggests that further bandwidth savings, well beyond those offered by today's foveated graphics approaches, could be enabled by exploiting such perceptual limitations with novel gaze-contingent hardware or software solutions. To the best of our knowledge, however, no perceptual model for eccentricity-dependent spatio-temporal aspects of human vision exists today, hampering the progress of such foveated graphics techniques.

The primary goal of our work is to experimentally measure user data and computationally fit models that adequately describe the eccentricity-dependent spatio-temporal aspects of human vision. Specifically, we design and conduct user studies with a custom, high-speed VR display that allow us to measure data to model
critical flicker fusion (CFF) in a spatially-modulated manner.
In this paper, we operationally define CFF as a measure of spatio-temporal flicker fusion thresholds. As such, and unlike current models of foveated vision, our model is unique in predicting what temporal information may be imperceptible for a certain eccentricity, spatial frequency, and luminance.

Using our model, we predict potential bandwidth savings of factors up to 3,500$\times$ over unprocessed visual information and 7$\times$ over existing foveated models that do not account for the temporal characteristics of human vision.

Specifically we make the following contributions:
\begin{itemize}	
	\item We design and conduct user studies to measure and validate eccentricity-dependent spatio-temporal flicker fusion thresholds with a custom display.
	\item We fit several variants of an analytic model to these data and also extrapolate the model beyond the space of our measurements using data provided in the literature, including an extension for varying luminance. 
	\item We analyze bandwidth considerations and demonstrate that our model may afford significant bandwidth savings for fo\-ve\-ated graphics.
\end{itemize}

\paragraph{Overview of Limitations.} The primary goals of this work are to develop a perceptual model and to demonstrate its potential benefits to foveated graphics. However, we are not proposing new foveated rendering algorithms or specific compression schemes that directly use this model.

\section{Related Work}
\label{sec:related}

\subsection{Perceptual Models}

The human visual system (HVS) is limited in its ability to sense variations in light intensity over space and time. Furthermore, these abilities vary across the visual field. The drop in spatial sensitivity with eccentricity has been well studied, attributed primarily to the drop in retinal ganglion cell densities~\cite{curcio1990topography} but also lens aberrations~\cite{shen2010peripheral, thibos1987vision}. 
This is often described by the spatial contrast sensitivity function (sCSF), defined as the inverse of the smallest contrast of sinusoidal grating that can be detected at each spatial frequency~\cite{robson1993contrast}. Spatial resolution, or acuity, of the visual system is then defined as the highest spatial frequency that can be seen, when contrast is at its maximum value of 1 or the log of the sCSF is zero.

On the contrary, temporal sensitivity has been observed to peak in the periphery, somewhere between $20-50^\circ$ eccentricity~\cite{hartmann1979peripheral, rovamo1984critical, tyler1987analysis}, attributed to faster cone cell responses in the mid visual field by Sinha~et~al.~\shortcite{sinha2017cellular}. Analogous to spatial sensitivity, temporal sensitivity is often described by the temporal contrast sensitivity function (tCSF) and detection at each temporal frequency. CFF thresholds then define the temporal resolution of the visual system. While Davis~et~al.~\shortcite{davis_humans_2015} recently showed that retinal image decomposition due to saccades in specific viewing conditions can elicit a flicker sensation beyond this limit, in our work we model only fixated sensitivities and highlight that further modeling would be needed to incorporate gaze-related effects (see Sec. 7).

While related, CFF and tCSF are used by vision scientists to quantify slightly different aspects of human vision. The CFF is considered to be a measure of conscious visual detection dependent on the temporal resolution of visual neurons, since at the CFF threshold, an identical flickering stimulus varies in percept from flickering to stable \cite{carmel2006conscious}. While contrast sensitivity is considered to be more of a measure of visual discrimination, as evidenced by the sinusoidal grating used in its measure requiring orientation recognition \cite{pelli2013measuring}. 

Several studies measure datapoints for these functions independently \cite{de1958research, van1967spatial, eisen2017evaluation}, across eccentricity~\cite{koenderink1978perimetry, hartmann1979peripheral}, as a function of luminance~\cite{koenderink1978perimetry3} and color~\cite{davis_humans_2015}, however few present quantitative models, possibly due to sparse sampling of the measured data. Additionally, data is often captured at luminances very different from typical VR displays. The Ferry--Porter~\cite{tyler1990analysis} and Granit--Harper \cite{rovamo1988critical} laws are exceptions to this, describing CFF as increasing linearly with log retinal illuminance and log stimulus area, respectively. Subsequent work by Tyler~et~al.~\shortcite{tyler1993eccentricity} showed that the Ferry-Porter law also extends to higher eccentricities. 
Koenderink~et~al.~\shortcite{koenderink1978visual} also derived a complex model for sCSF across eccentricity for arbitrary spatial patterns.

\begin{table}[t!]
\caption{\label{tab:models}Existing models of contrast sensitivity (CSF) and flicker fusion (CFF). Note that CSF models perception continuously across a range of conditions while CFF only models the limit of temporal perception at high temporal frequencies. Unlike ours, none of these models accounts for spatial and temporal variation as well as eccentricity and luminance.}
\begin{tabular}{lccccc}
\toprule
Model       & Spat. & Temp. & Eccentr. & Lum. & Prop. \\ 
\midrule
Tyler \shortcite{tyler1990analysis} & \xmark  & \cmark & \xmark & \cmark & CFF \\   
Kelly \shortcite{kelly_motion_1979} 				& \cmark  & \cmark & \xmark & \xmark & stCSF \\
Watson \shortcite{watson2016pyramid} 				& \cmark  & \cmark & \xmark & \cmark & stCSF \\
Watson \shortcite{watson2018field} 					& \cmark  & \xmark & \cmark & \cmark & sCSF \\
Ours        																& \cmark 	& \cmark & \cmark & \cmark & CFF \\
\bottomrule
\end{tabular}
\end{table}

However, spatial and temporal sensitivity are not independent. This relationship can be captured by varying both spatial and temporal frequency to obtain the spatio-temporal contrast sensitivity function (stCSF)~\cite{robson1966spatial}. While Kelly~\shortcite{kelly_motion_1979} was the first to fit an analytical function to stCSF data, this was limited to the fovea and a single luminance. Similarly, Watson~et~al.~\shortcite{watson2016pyramid} devised the pyramid of visibility, a simplified model that can be used if only higher frequencies are relevant. While also modeling luminance dependence, this model was again not applicable to higher eccentricities. Subsequent work saw the refitting of this model for higher eccentricity data but the original equation was simplified to only consider stationary content and the prediction of sCSF~\cite{watson2018field}.  

To the best of our knowledge, however, there is no unified model parameterized by all axes of interest, as illustrated in Table~\ref{tab:models}. With this work, we address this gap, measuring and fitting the first model that describes CFF in terms of spatial frequency, eccentricity and luminance. In particular, we measure CFF rather than stCSF to be more conservative in modeling the upper limit of human perception, for use in foveated graphics applications.

%%%%%%%%%%%%%%%%%%%%%%%%%%%%%%%%%%%%%%%%%%%%%%%%%%%%%%%%%%%%%%%%%%%%%%%%%%%%%%%%%%%%%%%%%%%%%%%%%%%%%%%%%%%%%%%%%%%%%%%%%%%%%%%%
\subsection{Foveated Graphics}

Foveated graphics encompasses a plethora of techniques that is enabled by eye tracking (see e.g.~\cite{Duchowski:2004,Koulieris:2019} for surveys on this topic). 
Knowing where the user fixates allows for manipulation of the visual data to improve perceptual realism~\cite{padmanaban2017optimizing,krajancich2020optimizing,mauderer2014depth}, or save bandwidth~\cite{arabadzhiyska2017saccade,albert2017latency}, by exploiting the drastically varying acuity of human vision over the retina. The most prominent approach that does the latter is perhaps foveated rendering, where images and videos are rendered, transmitted, or displayed with spatially varying resolutions without affecting the perceived image quality~\cite{geisler_real-time_1998,guenter_foveated_2012,patney_towards_2016,sun_perceptually-guided_2017,friston2019perceptual,Kaplanyan_2019_DeepFovea}. These methods primarily exploit spatial aspects of eccentricity-dependent human vision, but Tursun~et~al.~\shortcite{tursun2019luminance} recently expanded this concept by considering local luminance contrast across the retina, Xiao et al.~\shortcite{Lei_2020_neuralsuperres} additionally consider temporal coherence, and Kim et al.~\shortcite{kim2019foveated} showed hardware-enabled solutions for foveated displays. Related approaches also use gaze-contingent techniques to reduce bit-depth~\cite{mccarthy_sharp_2004} and shading or level-of-detail~\cite{luebke_perceptually_2001, murphy_gaze-contingent_2001, ohshima_gaze-directed_1996} in the periphery. All of these foveated graphics techniques primarily aim at saving bandwidth of the graphics system by reducing the number of vertices or fragments a graphics processing unit has to sample, raytrace, shade, or transmit to the display. To the best of our knowledge, none of these methods exploit the eccentricity-dependent temporal characteristics of human vision, perhaps because no existing model of human perception accounts for these in a principled manner (see Tab.~\ref{tab:models}). The goal of this paper is to develop such a model for eccentricity-dependent spatio-temporal flicker fusion that could enable significant bandwidth savings for all of the aforementioned techniques well beyond those enabled by existing eccentricity-dependent acuity models.

Many approaches in graphics, such as frame interpolation, temporal upsampling~\cite{Chen_2005_frame,scherzer_temporal_2012,didyk_perceptually-motivated_2010,denes_temporal_2019} and multi-frame rate rendering and display~\cite{springer2007multi}, do account for the limited temporal resolution of human vision. However, spatial and temporal sensitivities are not independent. Spatio-temporal video manipulation is common in foveated compression (e.g.,~\cite{ho2005practical,wang2003foveation}) 
while other studies have investigated how trading one for the other effects task performance, such as in first-person shooters~\cite{claypool_frame_2007,Claypool_2009}.
Denes~et~al.~\shortcite{denes_perceptual_2020} recently studied spatio-temporal resolution tradeoffs with perceptual models and applications to VR. However, none of these approaches rely on eccentricity-dependent models of spatio-temporal vision, which could further enhance their performance.

\section{Estimating Flicker Fusion Thresholds}
\label{sec:model_estimation}
To develop an eccentricity-dependent model of flicker fusion, we need a display that is capable of showing stimuli at a high framerate and over a wide FOV. In this section, we first describe a custom high-speed VR display that we built to support these requirements. We then proceed with a detailed discussion of the user study we conducted and the resulting values for eccentricity and spatial frequency--dependent CFF we estimated.

%%%%%%%%%%%%%%%%%%%%%%%%%%%%%%%%%%%%%%%%%%%%%%%%%%%%%%%%%%%%%%%%%%%%%%%%%%%%%%%
\subsection{Display Prototype}

\begin{figure}[t!]
  \centering
  \includegraphics[width=\linewidth]{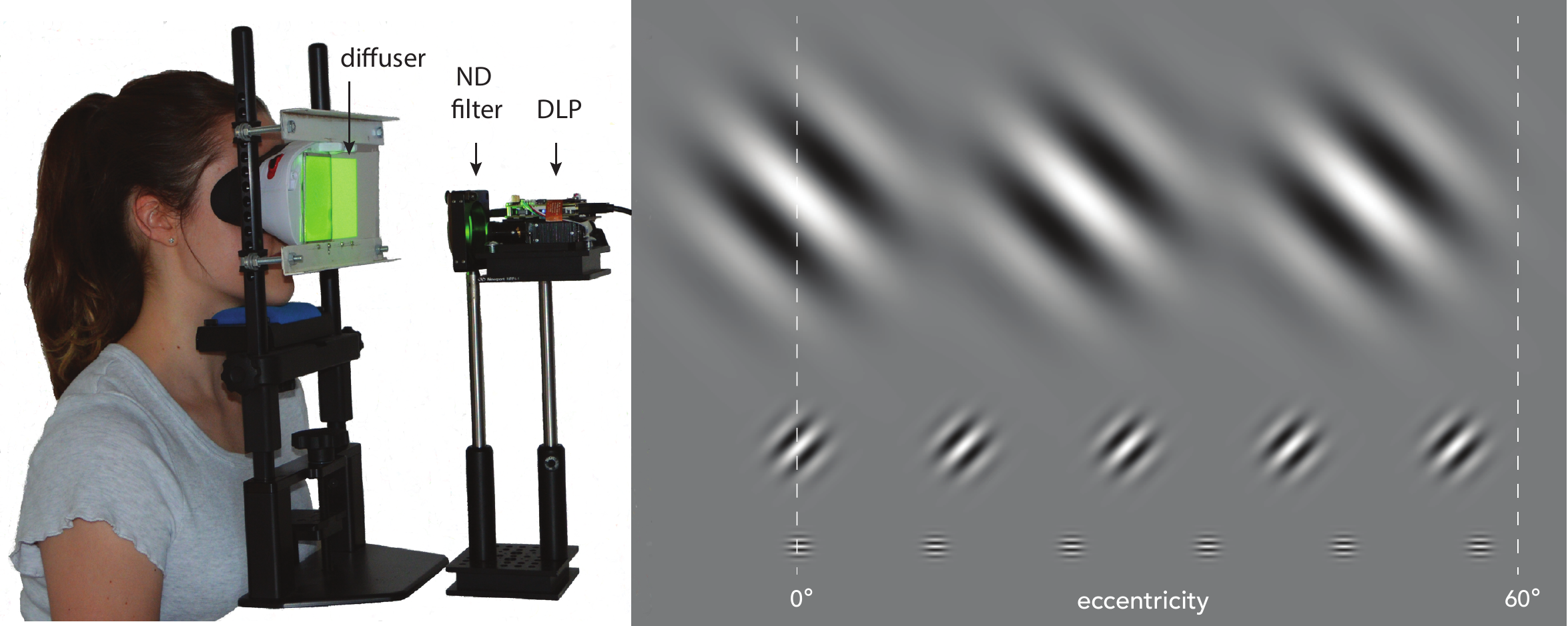}
  \caption{Our set-up for measuring user CFF thresholds. A photograph of a user in our custom VR display prototype is shown on the left. An ND filter is used to reduce the brightness of a DLP which projects onto a semi-transparent diffuser. On the right we show an illustration of the last 3 orders of test Gabor wavelets shown at a contrast of 1. Eccentricities were chosen to cover a FOV of $60^\circ$ at each scale. Orientation was chosen randomly from $45^\circ, 90^\circ, 135^\circ$ and $180^\circ$ at the point of beginning a QUEST staircase.}
  \label{fig:gabor_illustration}
\end{figure}

Our prototype display is designed in a near-eye display form factor to support a wide FOV. As shown in Figure~\ref{fig:gabor_illustration}(a), we removed the back panel of a View-Master Deluxe VR Viewer and mounted a semi-transparent optical diffuser (Edmund Optics \#47-679) instead of a display panel, which serves as a projection screen. This View-Master was fixed to an SR Research headrest to allow users to comfortably view stimuli for extended periods of time. To support a sufficiently high framerate, we opted for a Digital Light Projector (DLP) unit (Texas Instruments DLP3010EVM-LC Evaluation Board) that rear-projects images onto the diffuser towards the viewer. A neutral density (ND16) filter was placed in this light path to reduce the brightness to an eye safe level, measured to be $380$\,cd/m$^2$ at peak.

The DLP has a resolution of $1280 \times 720$, and a maximum frame rate of 1.5~kHz for 1-bit video, 360~Hz for 8-bit monochromatic video, or 120~Hz for 24-bit RGB video. 
We positioned the projector such that the image matched the size of the conventional View-Master display. Considering the magnification of the lenses, this display provides a pixel pitch of 0.1' (arc minutes) and a monocular FOV of $80^\circ$ horizontally and $87^\circ$ vertically. 

To display stimuli for our user study, we used the graphical user interface provided by Texas Instruments to program the DLP to the 360~Hz 8-bit grayscale mode, deemed sufficient for our measurements of CFF, which unlike CSF, does not require precise contrast tuning. 
The DLP was unable to support the inbuilt red, green and blue light-emitting diodes (LEDs) being on simultaneously, so we chose to use a single LED to minimize the possibility of artifacts from temporally multiplexing colors. Furthermore, since the HVS is most sensitive to mid-range wavelengths, the green LED (OSRAM: LE~CG~Q8WP) of peak wavelength 520~nm and a 100~nm full width at half maximum, was chosen so as to most conservatively measure the CFF thresholds. We used Python's PsychoPy toolbox~\cite{peirce2007psychopy} and a custom shader to stream frames to the display by encoding them into the required 24-bit RGB format that is sent to the DLP via HDMI. 

%%%%%%%%%%%%%%%%%%%%%%%%%%%%%%%%%%%%%%%%%%%%%%%%%%%%%%%%%%%%%%%%%%%%%%%%%%%%%%% 
\paragraph{Stimuli}

The flicker fusion model we wish to acquire could be parameterized by spatial frequency, rotation angle, eccentricity (i.e., distance from the fovea), direction from the fovea (i.e., temporal, nasal, etc.) and other parameters.
A naive approach may sample across all of these dimensions, but due to the fact that each datapoint needs to be recorded for multiple subjects and for many temporal frequencies per subject to determine the respective CFFs, this seems infeasible. Therefore, similar to several previous studies \cite{hartmann1979peripheral, allen1992visual, eisen2017evaluation}, we make the following assumptions to make data acquisition tractable:
\begin{enumerate}
\item The left and right eyes exhibit the same sensitivities and monocular and binocular viewing conditions are equivalent. Thus, we display the stimuli monocularly to the right eye by blocking the left side of the display.
\item Sensitivity is rotationally symmetric around the fovea, i.e. independent of nasal, temporal, superior, and inferior direction, thus being only a function of absolute distance from the fovea. It is therefore sufficient to measure stimuli only along the temporal direction starting from the fovea.
\item Sensitivity is orientation independent. Thus, the rotation angle of the test pattern is not significant.
\end{enumerate}
These assumptions allow us to reduce the sample space to only two dimensions: eccentricity $e$ and spatial frequency $\fs$.
Later we also analyze retinal illuminance $\lum$ as an additional factor.

Another fact to consider is that an eccentricity-dependent model that varies with spatial frequency must adhere to the uncertainty principle. That is, low spatial frequencies cannot be well localized in eccentricity. For example, the lowest spatial frequency of 0~cpd is a stimulus that is constant across the entire retina whereas very high spatial frequencies can be well localized in eccentricity. This behavior is appropriately modeled by wavelets. As such, we select our stimuli to be a set of 2D Gabor wavelets. As a complex sinusoid modulated by a Gaussian envelope, these wavelets are defined by the form:

\begin{align}
	g(\mathbf{x}, \mathbf{x_0}, \theta, \sigma, \fs) = \exp \left( \frac{-\lVert \mathbf{x}-\mathbf{x_0} \rVert^2}{2\sigma^2} \right) \cos \left(2\pi \fs \mathbf{x} \cdot [\cos \theta, \sin \theta]  \right),
\end{align}
where $\mathbf{x}$ denotes the spatial location on the display, $\mathbf{x_0}$ is the center of the wavelet, $\sigma$ is the standard deviation of the Gaussian in visual degrees, and $\fs$ and $\theta$ are the spatial frequency in cpd and angular orientation in degrees for the sinusoidal grating function.
We provide details of the conversion between pixels and eccentricity~$e$ in the Supplement.
Of note however, is that we use a scaled and shifted version of this form to be suitable for display, namely
$0.5 + 0.5\ g(\mathbf{x}, \mathbf{x_0}, \theta, \sigma, \fs)$, 
such that the pattern modulates between 0 and 1, with an average gray level of 0.5. 
The resulting stimulus exhibits three clearly visible peaks and smoothly blends into the uniform gray field which covers the entire field of view of our display and ends sharply at its edge with a dark background.

This choice of Gabor wavelet was motivated by many previous works in vision science, including the standard measurement procedure for the CSF~\cite{pelli2013measuring} (see Sec.~\ref{sec:related}), and image processing \cite{weldon1996efficient, kamarainen2006invariance}, where Gabor functions are frequently used for their resemblance to neural activations in human vision. 
For example, it has been shown that 2D Gabor functions are appropriate models of the transfer function of simple cells in the visual cortex of mammalian brains and thus mimicking the early layers of human visual perception~\cite{daugman1985uncertainty,marcelja1980mathematical}.

As a tradeoff between sampling the parameter space as densely as possible while keeping our user studies to a reasonable length, we converged on using 18 unique test stimuli, as listed in Table~\ref{tab:stimuli}. We sampled eccentricities ranging from $0^\circ$ to almost $60^\circ$, moving the fixation point into the nasal direction with increasing eccentricity, such that the target stimulus is affected as little as possible by the lens distortion. We chose not to utilize the full $80^\circ$ horizontal FOV of our display due to lens distortion becoming too severe in the last $10^\circ$. We chose to test 6 different spatial frequencies, 
with the highest being limited to 2~cpd due to the lack of a commercial display with both high enough spatial and temporal resolution. However it should be noted that the acuity of human vision is considerably higher; 60\,cpd based on peak cone density \cite{deering1998limits} and 40--50\,cpd based on empirical data \cite{thibos1987vision,robson1981probability,guenter_foveated_2012}. 
The Gaussian windows limiting the extent of the Gabor wavelets are scaled according to spatial frequency, i.e., $\sigma = 0.7/f_s$ such that each stimulus exhibits the same number of periods, defining the 6 wavelet orders. Finally, eccentricity values were chosen based on the radius of the wavelet order to uniformly sample the available eccentricity range, as illustrated in Figure~\ref{fig:gabor_illustration}(b).

\begin{table}
\caption{\label{tab:stimuli}Parameters of 18 test Gabor wavelets. We define 6 orders by spatial frequency (and radius) of stimuli. The number and eccentricity locations per order were chosen based on radius to uniformly sample the available eccentricity range. $f_s$: spatial frequency, $\sigma$: wavelet standard deviation, $e$: eccentricity.
(*) Note that in practice the extent was limited by our display FOV and $\fs = 0.0055$\,cpd is used for the analysis in Sec.~\ref{sec:model_fit}.
}
\begin{center}
\begin{tabular}{cccc} 
\toprule
 Order & $f_s$ (cpd) & $\sigma$ ($^\circ$) & $e$ ($^\circ$) \\
\midrule
 0 & 0.000(*) & inf(*) & \multicolumn{1}{l}{0.0} \\ 
 1 & 0.011 & 63.0 &  \multicolumn{1}{l}{0.0} \\ 
 2 & 0.041 & 17.2 &  \multicolumn{1}{l}{0.0, 19.2} \\ 
 3 & 0.154 & 4.6 &  \multicolumn{1}{l}{0.0, 24.5, 48.2} \\
4 & 0.571 & 1.2 &  \multicolumn{1}{l}{0.0, 14.8, 29.2, 42.7, 55.0}  \\ 
 5 & 2.000 & 0.5 &  \multicolumn{1}{l}{0.0, 12.3, 24.4, 35.9, 46.8, 56.8}  \\   
\bottomrule
\end{tabular}
\end{center}
\end{table}

The wavelets were temporally modulated by sinusoidally varying the contrast from $[-1, 1]$ and added to the background gray level of 0.5. In this way, at high temporal frequencies the Gabor wavelet would appear to fade into the background. The control stimulus was modulated at 180~Hz, which is far above the CFF observed for all of the conditions.

%%%%%%%%%%%%%%%%%%%%%%%%%%%%%%%%%%%%%%%%%%%%%%%%%%%%%%%%%%%%%%%%%%%%%%%%%%%%%%%
\subsection{User Study}

\paragraph{Participants}

Nine adults participated (age range 18--53, 4 female). Due to the demanding nature of our psychophysical experiment, only a few subjects were recruited, which is common for similar low-level psychophysics (see e.g. Patney et al.~\shortcite{patney_towards_2016}).
Furthermore, previous work measuring the CFF of 103 subjects found a low variance~\cite{romero2007value}, suggesting sufficiency of a small sample size.
All subjects in this and the subsequent experiment had normal or corrected-to-normal vision, no history of visual deficiency, and no color blindness, but were not tested for peripheral-specific abnormalities. All subjects gave informed consent. The research protocol was approved by the Institutional Review Board at the host institution.

\paragraph{Procedure}

To start the session, each subject was instructed to position their chin on the headrest such that several concentric circles centered on the right side of the display were minimally distorted (due to the lenses). The threshold for each Gabor wavelet was then estimated in a random order with a two-alternative forced-choice (2AFC) adaptive staircase designed using QUEST \cite{watson1983quest}. The orientation of each Gabor patch was chosen randomly at the beginning of each staircase from $0^\circ, 45^\circ, 90^\circ$ and $135^\circ$. At each step, the subject was shown a small ($1^\circ$) white cross for 1.5~s to indicate where they should fixate, followed by the test and control stimuli in a random order, each for 1~s. For stimuli at $0^\circ$ eccentricity, the fixation cross was removed after the initial display so as not to interfere with the pattern. The screen was momentarily blanked to a slightly darker gray level than the gray background to indicate stimuli switching. The subject was then asked to use a keyboard to indicate which of the two randomly ordered pattens (1 or 2) exhibited more flicker. The ability to replay any trial was also given via key press and the subjects were encouraged to take breaks at their convenience. Each of the 18 stimuli were tested once per user, taking approximately $90$~minutes to complete.

\paragraph{Results}

Mean CFF thresholds across subjects along with the standard error (vertical bars) and extent of the corresponding stimulus (horizontal bars) are shown in Figure~\ref{fig:model}. The table of measured values is available in the Supplemental Material.

The measured CFF values have a maximum above 90\,Hz. This relatively large magnitude can be explained by the Ferry--Porter law \cite{tyler1990analysis} and the high adaptation luminance of our display.
Similarly large values have previously been observed in corresponding conditions \cite{tyler1993eccentricity}.

As expected, the CFF reaches its maximum for the lowest $\fs$ values. This trend follows the Granit--Harper law predicting a linear increase of CFF with stimuli area~\cite{hartmann1979peripheral}.

For higher $\fs$ stimuli, we observe an increase of the CFF from the fovea towards a peak between $10^\circ$ and $30^\circ$ of eccentricity.
Similar trends have been observed by Hartmann\etal\shortcite{hartmann1979peripheral}, including the apparent shift of the peak position towards fovea with increasing $\fs$ and decreasing stimuli size.

Finally, our subjects had difficulty to detect flicker for the two largest eccentricity levels for the maximum $\fs = 2$\,cpd.
This is predictable as acuity drops close to or below this value for such extreme retinal displacements \cite{geisler_real-time_1998}.

\section{An Eccentricity-dependent Perceptual Model for Spatio-temporal Flicker Fusion}
\label{sec:model}

The measured CFFs establish an envelope of spatio-temporal flicker fusion thresholds at discretely sampled points within the resolution afforded by our display prototype. Practical applications, however, require these thresholds to be predicted continuously for arbitrary spatial frequencies and eccentricities. To this end, we develop a continuous eccentricity-dependent model for spatio-temporal flicker fusion that is fitted to our data. Moreover, we extrapolate this model to include spatial frequencies that are higher than those supported by our display by incorporating existing visual acuity data and we account for variable luminance adaptation levels by adapting the Ferry--Porter law~\cite{tyler1993eccentricity}.

\begin{figure*}[h]
  \centering
  \includegraphics[width=\linewidth]{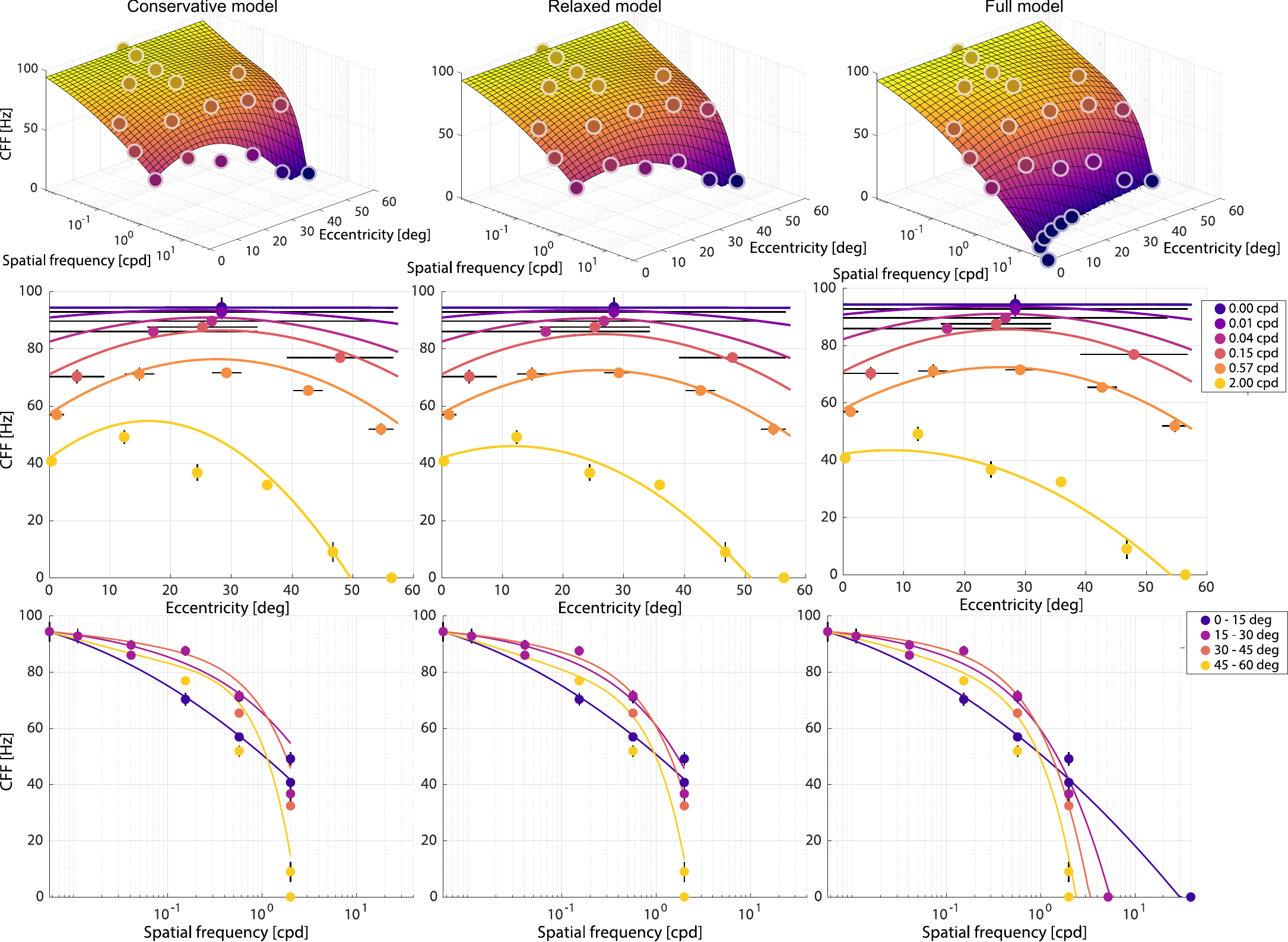}
  \caption{Three different parameter fits for our flicker fusion model $\cff(\ecc, \fs)$ (columns). 
	Orthographic views of the eccentricity--CFF (second row) and spatial frequency--CFF (third row) planes show subject-averaged measured points along with their variances (standard error of means, vertical bars). 
  The horizontal bars in the eccentricity dimension denote spatial extents of respective stimuli.
  The curves represent corresponding cross sections of the fitted models.
  }
  \label{fig:model}
\end{figure*}

\subsection{Model Fitting}
\label{sec:model_fit}

Each of our measured data points is parametrized by its spatial frequency $\fs$, eccentricity $\ecc$, and CFF value averaged over all subjects.
Furthermore, it is associated with a localization uncertainty determined by the radius of its stimulus $\radius$.
In our design, $\radius$ is a function of $\fs$ and, for 13.5\% peak contrast cut-off, we define $\radius = 2 \sigma$ where $\sigma = 0.7/f_s$ to be the standard deviation of our Gabor patches.
\begin{table}[b]
\caption{\label{tbl:model_parameters}Parameters $p_{0 \ldots 9}$ for our model fitted for conservative (cons.) and relaxed (rel.) assumptions as well as the full modeled extended using acuity data. The degrees-of-freedom adjusted R$^2$ shows the fit quality.}
\footnotesize
\begin{tabular}{@{}lll@{}}
\toprule
Model & \multicolumn{1}{c}{Parameters} & \multicolumn{1}{c}{$R^2$} \\ 
            & \multicolumn{1}{c}{$ \ \ p_0 \ \ \ $ $ \ \ p_1 \ \ $ $ \ \ p_2 \ \ \ $ $ \ \ p_3 \ \ \ $ $ \ \ \ p_4 \ \ $ $ \ \ \ \ \ p_5 \ \ \  $ $\ \ \ \ \ \ p_6 \ \ \ \ \ $ $\ \ \ \ p_7 \ \ \ \ \ $ $\ \ \ \ \ p_8 \ \ \ $  $ \ \ \ p_9 \ \ \ $} & \multicolumn{1}{c}{} \\ \midrule
Cons. & -4.08, -10.1, 94.4, 0.0484, -0.280, 0.431, -0.00140, 0.00679, -0.00912, 1.56 & 0.889 \\  
Rel.      & -4.06, -10.1, 94.3, 0.0464, -0.282, 0.430, -0.00129, 0.00672, -0.00929, 1.58 & 0.938 \\
Full         & -4.06, -10.1, 94.3, 0.0440, -0.281, 0.435, -0.00111, 0.00613, -0.00877, 1.58 & 0.938 \\ \bottomrule
\end{tabular}
\end{table}

We formulate our model as
\begin{align}
\cff(\ecc, \fs) &= \max(0, p_0 \tau(\fs)^2 + p_1 \tau(\fs) + p_2 \\
				&+ (p_3 \tau(\fs)^2 + p_4 \tau(\fs) + p_5) \cdot \zeta(\fs) \ecc \nonumber \\
				&+ (p_6 \tau(\fs)^2 + p_7 \tau(\fs) + p_8) \cdot \zeta(\fs) \ecc^2) \nonumber \\
\zeta(\fs) &= \exp(p_9 \tau(\fs)) - 1 \nonumber \\    
\tau(\fs) &= \max(\log_{10}{\fs} - \log_{10}{f_{s_{0}}}, 0), \nonumber
\label{eq:cff}
\end{align}

where $\mathbf{p} = [p_0, \ldots, p_9] \in \real^{10}$ are the model parameters (see Table~{\ref{tbl:model_parameters}}),
$\zeta(\fs)$ restricts eccentricity effects for small $\fs$ and $\tau(\fs)$ offsets logarithmic $\fs$ relative to our constant function cut-off.

We build on three domain-specific observations to find a continuous CFF model $\cff(\ecc, \fs): \real^2\to\real$ that fits our measurements.

First, both our measurements and prior work indicate that the peak CFF is located in periphery, typically between $20^\circ$ and $50^\circ$ of eccentricity \cite{hartmann1979peripheral,rovamo1984critical,tyler1987analysis}.
For both fovea and far periphery the CFF drops again forming a convex shape which we model as a quadratic function of $\ecc$.

Second, because the stimuli with very low $\fs$ are not spatially localized, their CFF does not vary with $\ecc$.
Consequently, we enforce the dependency on $\ecc$ to converge to a constant function for any $\fs$ below $f_{s_{0}} = 0.0055$~cpd.
This corresponds to half reciprocal of the full-screen stimuli visual field coverage given our display dimensions.

Finally, following common practices in modeling the effect of spatial frequencies on visual effects, such as contrast \cite{koenderink1978perimetry2} or disparity sensitivities \cite{bradshaw1999sensitivity}, we fit the model for logarithmic $\fs$.

Before parameter optimization, we need to consider the effect of eccentricity uncertainty.
The subjects in our study detected flicker regardless of its location within the stimuli extent $\extent = [\ecc \pm \radius]$\,deg. 
Therefore, $\cff$ achieves its maximum within $\extent$ and is upper-bounded by our measured flicker frequency $\pt \in \real$.
At the same time, nothing can be claimed about the variation of $\cff$ within $\extent$ and therefore, in absence of further evidence, a conservative model has to assume that $\cff$ is not lower than $\pt$ within $\extent$.
These two considerations delimit a piece-wise constant $\cff$.
In practice, based on previous work \cite{hartmann1979peripheral,tyler1987analysis} it is reasonable to assume that $\cff$ follows a smooth trend over the retina and its value is lower than $\pt$ value at almost all eccentricities within $\extent$.

Consequently, in Table~{\ref{tbl:model_parameters}} we provide two different fits for the parameters.
Our conservative model strictly follows the restrictions from the measurement and tends to overestimate the range of visible flicker frequencies which prevents discarding potentially visible signal.
Alternatively, our relaxed model follows the smoothness assumption and applies the measured values as upper bound.

To fit the parameters, we used the Adam solver in PyTorch initialized by the Levenberg--Marquardt algorithm and we minimized the mean-square prediction error over all extents $\extent$. 
The additional constraints were implemented as soft linear penalties.
To leverage data points with immeasurable CFF values, we additionally force $\cff(\ecc, \fs) = 0$ at these points.
This encodes imperceptibility of their flicker at any temporal frequency.

Figure~\ref{fig:model} shows that the fitted $\cff$ represents the expected effects well. The eccentricity curves (row~2) flatten for low $\fs$ and their peaks shift to lower $\ecc$ for large $\fs$. The conservative fit generally yields larger CFF predictions though it does not strictly adhere to the stimuli extents due to other constraints.

\subsection{Extension for High Spatial Frequencies}
Due to technical constraints, the highest $\fs$ measured was 2\,cpd. At the same time, the acuity of human vision has an upper limit of 60\,cpd based on peak cone density \cite{deering1998limits} and 40--50\,cpd based on empirical data \cite{thibos1987vision,robson1981probability,guenter_foveated_2012}. To minimize this gap and to generalize our model to other display designs we extrapolate the CFF at higher spatial frequencies using existing models of spatial acuity.

For this purpose, we utilize the acuity model of Geisler and Perry~\shortcite{geisler_real-time_1998}.
It predicts acuity limit $\acuity$ for $\ecc$ as
\begin{align}
\acuity(\ecc) = \ln(64) \frac{2.3}{0.106 \cdot (\ecc + 2.3)},
\end{align}
with parameters fitted to measurements of Robson and Graham~\shortcite{robson1981probability}.
Their study of pattern detection rather than resolution is well aligned with our own study design and conservative visual performance assessment.
Similarly, their bright adaptation luminance of 500\,cd/m$^2$ is also close to our display.

$\acuity(\ecc)$ predicts limit of spatial perception. 
We reason that at this absolute limit flicker is not detectable and, therefore, the CFF is not defined.
We represent this situation by zero CFF values in the same way as for imperceptible stimuli in our study and force our model to satisfy $\cff(\ecc, \acuity(\ecc)) = 0$.

In combination with our relaxed constraints we obtain our final full model as shown in Figure~\ref{fig:model}.
It follows the same trends as our original model within the bounds of our measurement space and intersects the zero plane at the projection of $\acuity(\ecc)$.

%%%%%%%%%%%%%%%%%%%%%%%%%%%%%%%%%%%%%%%%%%%%%%%%%%%%%%%%%%%%%%%%%%%%%%%%%%%%%%%%%
\subsection{Adaptation luminance}

Our experiments were conducted at half of our display peak luminance $\lumcd = 380$\,cd/m$^2$. 
This is relatively bright compared to the 50--200\,cd/m$^2$ luminance setting of common VR systems \cite{mehrfard2019comparative}.
Consequently, our estimates of the CFF are conservative because the Ferry--Porter law predicts the CFF to increase linearly with logarithmic levels of retinal illuminance \cite{tyler1993eccentricity}.
While the linear relationship is known, the actual slope and intercept varies with retinal eccentricity \cite{tyler1993eccentricity}.
For this reason, we measured selected points from our main experiment for two other display luminance levels.

Four of the subjects from experiment~1 performed the same procedure for a subset of conditions with a display modified first with one and then two additional ND8 filters yielding effective luminance values of 23.9 and 3.0\,cd/m$^2$.
We then applied a formula of Stanley and Davies~\shortcite{stanley1995effect} to compute the pupil diameter under these conditions as 
\begin{align}
d(\lumcd) &= 7.75 - 5.75 \left(\frac{(\lumcd a/846)^{0.41}}{(\lumcd a/846)^{0.41}+2}\right), 
\end{align}
where $a = 80\times87 = 6960$\,deg$^2$ is the adapting area of our display.
This allows us to derive corresponding retinal illuminance levels for our experiments using $\lum(\lumcd) = \pi d(\lumcd)^2/4 \cdot \lumcd$ as 67.3, 321 and 1488\,Td and obtain a linear transformation of our original model $\cff(\ecc,\fs)$ to account for $\lumcd$ with an eccentricity-dependent slope as
\begin{align}
\hat{\cff}(\ecc,\fs,\lumcd) &= (s(\ecc,\fs)\cdot(\log_{10}(\lum(\lumcd)/\lum_0)) + 1) \cff(\ecc,\fs) \\
s(\ecc,\fs) &= \zeta(\fs)(q_0 \ecc^2 + q_1 \ecc) + q_2
\end{align}
where $\lum_0 = 1488$~Td is our reference retinal illuminance,
$\zeta(\fs)$ encodes localization uncertainty for low $\fs$ as in Equation~\ref{eq:cff} and $\mathbf{q} = [5.71\cdot 10^{-6}, -1.78\cdot 10^{-4}, 0.204]$ are parameters obtained by a fit with our full model.

\begin{figure}[t!]
  \centering
  \includegraphics[width=\linewidth]{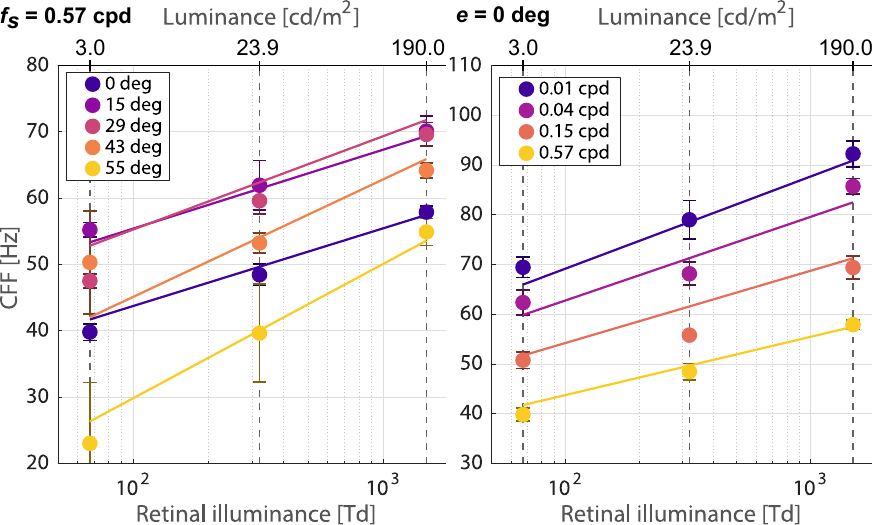}
  \caption{
  Luminance scaling of our full model fitted for $\lum_0 = 1488$\,Td. 
  The points represent mean measured CFF values and the lines the prediction of the transformed model.
  Vertical bars are standard errors.
  The left plot shows varying slopes across $\ecc$ (with $\fs = 0.57$\,cpd). 
  The right plot shows varying intercepts across $\fs$ (with $\ecc = 0$\,deg). 
  }
  	\label{fig:luminance}
\end{figure}

Figure~\ref{fig:luminance} shows that this eccentricity-driven model of Ferry--Porter luminance scaling models not only the slope variation over the retina but also the sensitivity difference over range of $\fs$ well (degree-of-freedom-adjusted R$^2$ = 0.950).

\section{Experimental Validation}
\label{sec:validation}

Our eccentricity-dependent spatio-temporal model is unique in that it allows us to predict what temporal information may be imperceptible for a certain eccentricity and spatial frequency. One possible application for such a model is in the development of new perceptual video quality assessment metrics (VQMs). Used to guide the development of different video codecs, encoders, encoding settings, or transmission variants, such metrics aim to predict subjective video quality as experienced by users. 
While it is commonly known that many existing metrics, such as peak signal-to-noise ratio (PSNR) and structural similarity index metric (SSIM), do not capture many eccentricity-dependent spatio-temporal aspects of human vision well, in this section we discuss a user study we conducted which shows that our model could help better differentiate perceivable and non-perceivable spatio-temporal artifacts. Furthermore, we also use the study to test the fit derived in the previous section, showing that our model makes valid predictions for different users and points beyond those measured in our first user study.

\begin{figure}[t!]
  \centering
  \includegraphics[width=\linewidth]{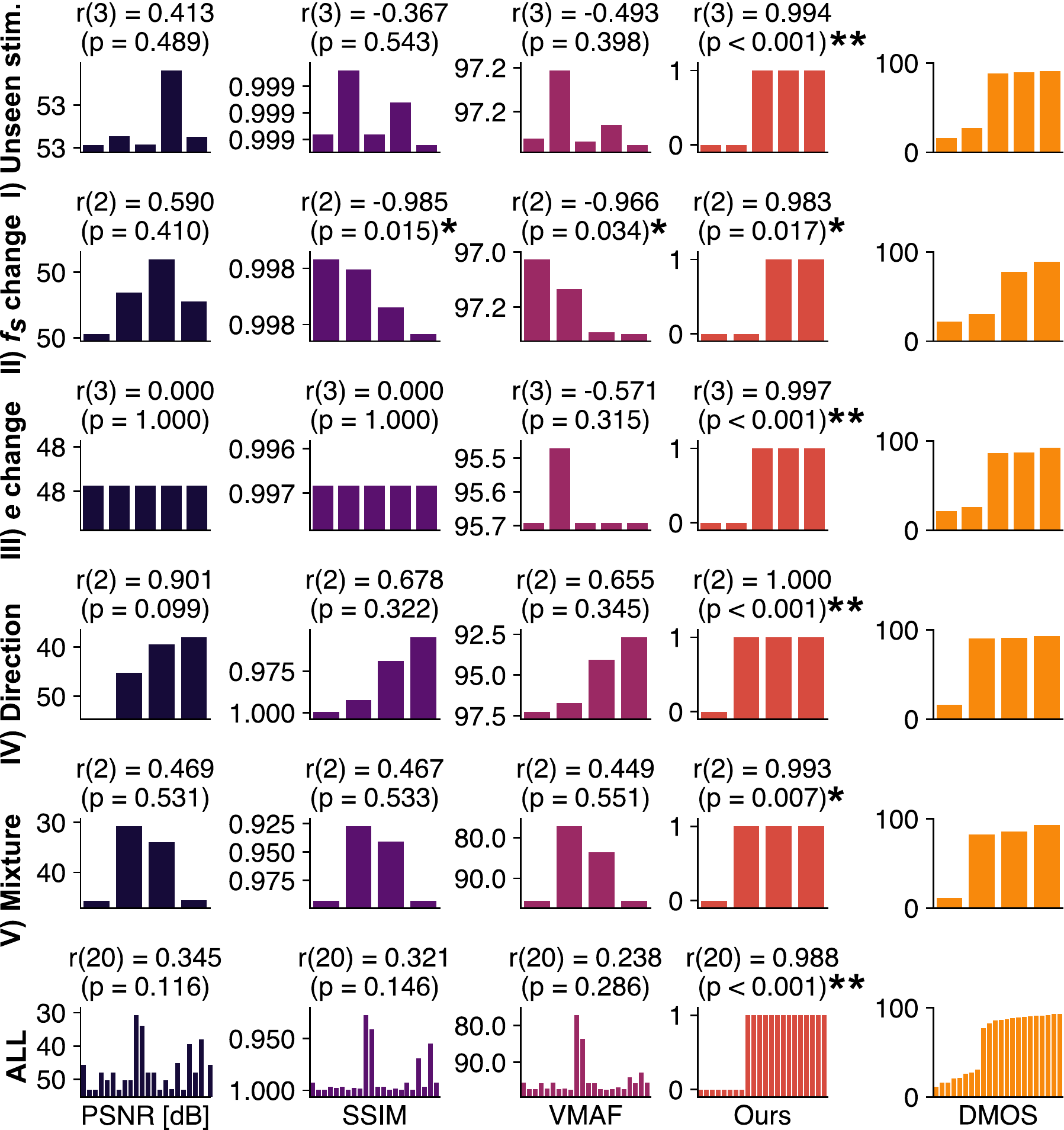}
  \caption{Results of our validation study. 
  Each row represents one of the 5 tested effects along with a cumulative plot in the last row.
  Results of different metrics (vertical columns) are compared to DMOS (right column).
  Each bar represents the score for one stimulus rescaled so that a higher bar indicates a better quality.
PLCC of each metric and DMOS are shown along the p-values (t-test).
(*) and (**) mark statistical significance at $p=0.05$ and $p=0.001$ levels respectively.
  }
  	\label{fig:validation}
\end{figure}

The study was conducted using our custom high-speed VR display with 18 participants, 13 of which did not participate in the previous user study. Users were presented with videos, made up of a single image frame perturbed by Gabor wavelet(s) such that when modulated at a frequency above the CFF they become indistinguishable from the background image. Twenty-two unique videos were tested, where the user was asked to rank the quality of the video from 1 (``bad'') to 5 (``excellent''). 
We then calculate the difference mean opinion score (DMOS) per stimulus, as described in detail by Seshadrinathan~et~al~\shortcite{seshadrinathan2010study}, and 3 VQMs, namely PSNR, SSIM and one of the most influential metrics used today for traditional contents: the Video Multimethod Assessment Fusion (VMAF) metric developed by Netflix \cite{li2016toward}. The results of which are shown in Figure~\ref{fig:validation}.
We acknowledge that neither of these metrics has been designed to specifically capture the effects we measure. This often results in uniform scores across the stimuli range (see e.g. Figure~\ref{fig:validation} III). We include them into our comparison to illustrate these limits and the need for future research in this direction.

The last row shows a comparison of all measured DMOS values with each of the standard VQMs and our own flicker quality predictor.
This is a simple binary metric that assigns videos outside of the CFF volume where flicker is invisible a good label while it assigns a bad label to others.
Our full $\cff(\ecc, \fs)$ model was used for this purpose.
We computed Pearson Linear Correlation Coefficients (PLCC) between DMOS and each of the metrics while taking into account their semantics.
Only our method exhibits statistically significant correlation with the user scores ($r(20) = 0.988$, $p < 0.001$ from a two-tail t-test).

Further, we broke our stimuli into 5 groups to test the ability of our model to predict more granular effects.

\begin{enumerate}
\item \emph{Unseen stimulus.} We select $\ecc = 30^\circ$ and $\fs = 1.00$~cpd, a configuration not used to fit the model, and show that users rate the video to be of poorer quality (lower DMOS) when it is modulated at a frequency lower than the predicted CFF (see panel I in Fig.~\ref{fig:validation}). 
\item \emph{$f_s$ dependence.} We test whether our model captures spatio-temporal dependence. Lowering $\fs$ with fixed radius makes the flicker visible, despite not changing $\ft$ (panel II).
\item \emph{$e$ change.} By moving the fixation point towards the outer edges of the screen, thereby increasing $\ecc$, we show that our model captures and predicts the higher temporal sensitivity of the mid-periphery.
 Furthermore, by moving the fixation point in both nasal and temporal directions, we validate the assumed symmetry of the CFF. As a result, users were able to notice flicker that was previously imperceptible in the fovea (panel III).
\item \emph{Direction independence.} We confirm our assumption that sensitivity is approximately rotationally symmetric around the fovea. A single visible stimulus from Group 1 was also added for the analysis as a control (panel IV).
\item \emph{Mixed stimuli.} We show that the model predictions hold up even with concurrent viewing of several stimuli of different sizes, eccentricities and spatial frequencies (panel V).
\end{enumerate}

The perturbation applied to the videos in Groups 1–3 only varied in modulation frequency or eccentricity with respect to a moving gaze position. In response, the dependency of other metrics on $\ft$ is on the level of noise, since changes related to variation of $\ft$ or $\fs$  of a fixed-size wavelet average out over several frames, and moving the gaze position does not change the frames at all. Our metric is able to capture the perceptual effect, maintaining significant correlation with DMOS ($r(3) = 0.994$, $p = 0.001$, $r(2) = 0.983$, $p < 0.05$ and $r(3) = 0.977$, $p < 0.001$ respectively).

Groups 4 and 5 were designed to be accumulating sets of Gabor wavelets, but all with parameters that our model predicts as imperceptible. We observe that VQMs drop with increasing total distortion area, while the DMOS scores stay relatively constant and significantly correlated with our metric ($r(2) = 1.000$, $p < 0.001$ and $r(3) = 0.993$, $p < 0.01$ respectively). These results indicate that existing compression schemes may be able to utilize more degrees of freedom to achieve higher compression, without seeing a drop in the typical metrics used to track perceived visual quality. The exact list of chosen parameters, along with the CFFs predicted by our model are listed in the supplemental material.

In conclusion, the study shows that our metric predicts visibility of temporal flicker for independently varying spatial frequencies and retinal eccentricities.
While existing image and video quality metrics are important for predicting other visual artifacts, our method is a novel addition in this space that significantly improves perceptual validity of quality predictions for temporally varying content. In this way, we also show that our model may enable existing compression schemes to utilize more degrees of freedom to achieve higher compression, without compromising several conventional VQMs.

\section{Analyzing Bandwidth Considerations}
\label{sec:analysis}
Our eccentricity-dependent spatio-temporal CFF model defines the gamut of visual signals perceivable to human vision.
Hence any signal outside of this gamut can be removed to save bandwidth in computation or data transmission.
In this section we provide a theoretical analysis of the compression gain factors this model may potentially enable for foveated graphics applications when used to allocate resources such as bandwidth.
In practice, developing foveated rendering and compression algorithms holds other nuanced challenges and thus the reported gains represent an upper bound.

To efficiently apply our model, a video signal should be described by a decomposition into spatial frequencies, retinal eccentricities, and temporal frequencies.
We consider discrete wavelet decompositions (DWT) as a suitable candidate because these naturally decompose a signal into eccentricity-localized spatio-temporal frequency bands. 
Additionally, the multitude of filter scales in the hierarchical decomposition closely resembles scale orders in our study stimuli.

For the purpose of this thought experiment, we use a biorthogonal Haar wavelet, which, for a signal of length $N$, results in $\log_2(N)$ hierarchical planes of recursively halving lengths.
The total number of resulting coefficients is the same as the input size.
From there our baseline is retaining the entire original set of coefficients yielding compression gain of 1.

\begin{figure}[t!]
  \centering
  \includegraphics[width=\linewidth]{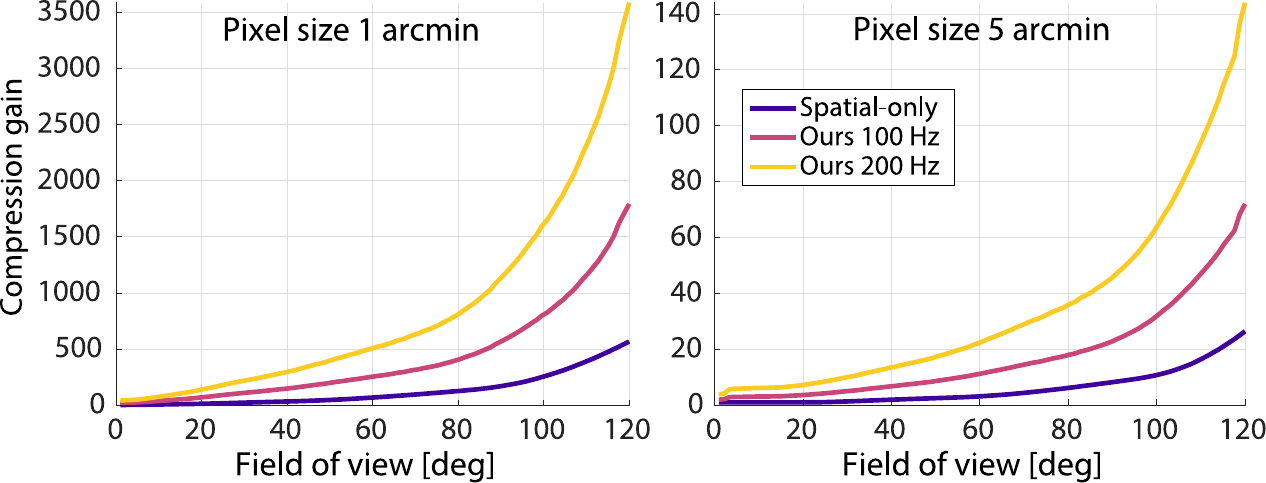}
  \caption{Compression gain analysis as a fraction of original and retained decomposition coefficients depending on the screen pixel density (left and right plots) and the covered visual field (horizontal axis). Note the difference in the gain axes scales.}
  \label{fig:compression_analysis}
\end{figure}

For traditional spatial-only foveation, we process each frame independently and after a 2D DWT we remove coefficients outside of the acuity limit.
We compute eccentricity assuming gaze in the center of the screen and reject coefficients for which $\cff(\ecc, \fs) = 0$.
From our model definition, such signals cannot be perceived regardless of $\ft$ as they lie outside of the vision acuity gamut.
The resulting compression gain compared to the baseline can be seen for various display configurations in Figure~\ref{fig:compression_analysis}.

Next, for our model we follow the same procedure but we additionally decompose each coefficient of the spatial DWT using 1D temporal DWT. This yields an additional set of temporal coefficients $\ft$ and we discard all values with $\cff(\ecc, \fs) < \ft$.

For this experiment, we assume a screen with the same peak luminance of 380\,cd/m$^2$ as our display prototype and a pixel density of 60 pixel per degree (peak $\fs = 30$\,cpd) for approximately retinal display or a fifth of that for a display closer to existing VR systems.
The tested aspect ratio of 165:135 approximates the human binocular visual field~\cite{Ruch:1960}.
We consider displays with a maximum framerate of 100\,Hz (peak reproducible $f_t = 50$\,Hz) and 200\,Hz, capable of reaching the maximum CFF in our model. 
The conventional spatial-only foveation algorithm is not affected by this choice.

Figure~\ref{fig:compression_analysis} shows that this significantly improves compression for both small and large field-of-view displays regardless of pixel density.
This is because the spatial-only compression needs to retain all temporal coefficients in order to prevent the worst-scenario flicker at $\ecc$ and $\fs$ with highest CFF.
Therefore, it only discards signal components in the horizontal $\ecc \times \fs$ plane.
On the other hand, our scheme allows to also discard signal in the third temporal dimension closely copying the shape of the $\cff$ which allows to reduce retained $\ft$ for both fovea and far periphery in particular for high $\fs$ coefficients which require the largest bandwidth.

\section{Discussion}
\label{sec:discussion}
The experimental data we measure and the models we fit to them further our understanding of human perception and lay the foundation of future spatio-temporal foveated graphics techniques. Yet, several important questions remain to be discussed. 

\paragraph{Limitations and Future Work}

First, our data and model work with CFF, not the CSF. CFF models the temporal thresholds at which a visual stimulus is predicted to become (in)visible. Unlike the CSF, however, this binary value of visible/invisible does not model relative sensitivity. This makes it not straightforward to apply the CFF directly as an error metric, for example to optimize foveated rendering or compression algorithms. Obtaining an eccentricity-dependent model for the spatio-temporal CSF is an important goal for future work, yet it requires a dense sampling of the full three-dimensional space spanned by eccentricity, spatial frequency, and time with user studies. The CFF, on the other hand only requires us to determine a single threshold for the two-dimensional space of eccentricity and spatial frequency. Considering that obtaining all 18 sampling points in our 2D space for a single user already takes about 90 minutes, motivates the development of a more scalable approach to obtaining the required data in the future. 

Although we validate the linear trends of luminance-dependent behavior of the CFF predicted by the Ferry--Porter law, it would be desirable to record more users for larger luminance ranges and more densely spaced points in the 2D sampling space. Again, this would require a significantly larger amount of user studies, which seems outside the scope of this work. Similarly, we extrapolate our model for $\fs$ above the 2\,cpd limit of our display. Future work may utilize advancements in display technology or additional optics to confirm the model extension. We also map the spatio-temporal thresholds only for monocular viewing conditions. Some studies have suggested that such visual constraint lowers the measured CFF in comparison to binocular viewing~\cite{eisen2020dichoptic,ali1991critical}. Further work is required to confirm the significance of such an effect in the context of displays.

It should also be noted that our model assumes a specific fixation point. We did not account for the dynamics of ocular motion, which may be interesting for future explorations. 
By simplifying temporal perception to be equivalent to the temporal resolution of the visual system we also do not account for effects caused by our pattern sensitive system, such as our ability to detect spatio-temporal changes like local movements or deformations. 
Similarly, we do not model the effects of crowding in the periphery, which has been shown to dominate spatial resolution in pattern recognition~ \cite{rosenholtz2016capabilities}. Finally, all of our data is measured for the green color channel of our display and our model assumes rotational invariance as well as orientation independence of the stimulus. A more nuanced model that studies variations in these dimensions may be valuable future work. 
All these considerations should be taken into account when developing practical compression algorithms, which is a task beyond the scope of this paper.

Finally, we validate that existing error metrics, such as PSNR, SSIM, and VMAF, do not adequately model the temporal aspects of human vision. The correlation between our CFF data and the DMOS of human observers is significantly stronger, motivating error metrics that better model these temporal aspects. Yet, we do not develop such an error metric nor do we propose practical foveated compression or rendering schemes that directly exploit our CFF data. These investigations are directions of future research.

\paragraph{Conclusion}

At the convergence of applied vision science, computer graphics, and wearable computing systems design, foveated graphics techniques will play an increasingly important role in emerging augmented and virtual reality systems. With our work, we hope to contribute a valuable foundation to this field that helps spur follow-on work exploiting the particular characteristics of human vision we model.

\begin{acks}
B.K. was supported by a Stanford Knight-Hennessy Fellowship. G.W. was supported by an Okawa Research Grant, a Sloan Fellowship, and a PECASE by the ARO. Other funding for the project was provided by NSF (award numbers 1553333 and 1839974). 
The authors would also like to thank Brian Wandell, Anthony Norcia, and Joyce Farrell for advising on temporal mechanisms of the human visual system, Keith Winstein for expertise used in the development of the validation study, Darryl Krajancich for constructing the apparatus for our custom VR display, and Yifan (Evan) Peng for assisting with the measurement of the display luminance.
\end{acks}

% Bibliography
\bibliographystyle{ACM-Reference-Format}
\bibliography{ms}

% Appendix
%\appendix

\end{document}